\documentclass[english]{cccconf}
\usepackage[comma,numbers,square,sort&compress]{natbib}
\usepackage{epstopdf}

\usepackage{tikz}
\graphicspath{{images/}}
\usepackage{amsmath}
\usepackage{amsmath,bm}
\usepackage{latexsym, amssymb}
\usepackage{graphicx}
\usepackage{amsmath, amsbsy}
\usepackage{amsopn, amstext}
\usepackage{ifpdf,hyperref}
\usepackage{cancel, color}
\usepackage{epstopdf}

\def\J{{\bf 1}}

\DeclareMathOperator{\sgn}{sgn}
\DeclareMathOperator{\Det}{Det}
\DeclareMathOperator{\ddet}{ddet}
\DeclareMathOperator{\cdet}{cdet}

\DeclareMathOperator{\lcm}{lcm}

\DeclareMathOperator{\GL}{GL}

\def\ra{\rightarrow}

\def\a{\alpha}
\def\b{\beta}
\def\d{\delta}

\def\D{\Delta}

\def\0{{\bf 0}}

\newcommand{\R}{{\mathbb R}}
\newcommand{\C}{{\mathbb C}}

\newcommand{\F}{{\mathbb F}}

\def\dsum{\mathop{\sum}\limits}
\def\thefootnote{*}

\newtheorem{thm}{Theorem}[section]
\newtheorem{dfn}[thm]{Definition}
\newtheorem{prp}[thm]{Proposition}
\newtheorem{exa}[thm]{Example}

\newtheorem{cor}[thm]{Corollary}
\newtheorem{rem}[thm]{Remark}

\begin{document}
	
\title{Semi-Tensor Product of Hypermatrices with Application to Compound Hypermatrices}

	
	\author{Daizhan Cheng\aref{amss}, Xiao Zhang\aref{amss,ncmis}, Zhengping Ji \aref{amss,ucas}}
	
	
	%
	
	\affiliation[amss]{Key Laboratory of Systems and Control, Academy of Mathematics and Systems Science,
		Chinese Academy of Sciences, Beijing 100190, P.~R.~China}
	
	\affiliation[ucas]{School of Mathematical Sciences, University of Chinese Academy of Sciences, Beijing 100049, P.~R.~China}
	
	\affiliation[ncmis]{National Center for Mathematics and Interdisciplinary Sciences, Chinese Academy of Sciences, Beijing 100190, P.~R.~China\email{~dcheng@iss.ac.cn~xiaozhang@amss.ac.cn;~jizhengping@amss.ac.cn}}
	
	\maketitle
	
	\begin{abstract}
		The semi-tensor product (STP) of matrices is extended to the STP of hypermatrices. Some basic properties of the STP of matrices are extended to the STP of hypermatrices. The hyperdeterminant of hypersquares is introduced. Some algebraic and geometric structures of matrices are extended to hypermatrices. Then the compound hypermatrix is proposed. The STP of hypermatrix is used to compound hypermatrix. Basic properties are proved to be available for compound hypermatrix.
	\end{abstract}
	
	\keywords{Semi-tensor product,  $d$-hypermatrix, general linear group of hypermatrices, compound hypermatrices}
	
	\footnotetext{This work is supported partly by the National Natural Science Foundation of China (NSFC) under Grants 62073315 and 61733018.}
	
\section{Introduction}
The last two decades have witnessed the rapid development of the semi-tensor product (STP) of matrices \cite{che11,che12}. In particular, it has been applied to study Boolean networks and finite valued networks (see survey papers \cite{for16,li18,lu17,muh16}); finite games (see survey paper \cite{che21}); finite automata (see survey paper \cite{yan22}); dimension-varying systems \cite{che19,che19b}, etc. The STP of two matrices is defined as follows:

\begin{dfn}\cite{che11}\label{d1.1} Let $A\in \F^{m\times n}$ and $B\in \F^{p\times q}$ and $t=\lcm(n,p)$. The STP of $A$ and $B$ is defined by
\begin{align}\label{1.1}
A\ltimes B:=\left(A\otimes I_{t/n}\right)\left(B\otimes I_{t/p}\right).
\end{align}
\end{dfn}

Some basic properties of STP are as follows:

\begin{prp}\label{p1.2}
\begin{itemize}
\item[(i)] (Linearity)
\begin{align}\label{1.2}
\begin{array}{l}
A\ltimes (\a B+\b C)=\a A\ltimes B+\b A\ltimes C,\\
(\a B+\b C)\ltimes A=\a B\ltimes A+\b C\ltimes A,\quad \a,\b\in \F.\\
\end{array}
\end{align}
\item[(ii)] (Associativity)
\begin{align}\label{1.3}
A\ltimes (B\ltimes C)=(A\ltimes B)\ltimes C.
\end{align}
\end{itemize}
\end{prp}

\begin{prp}\label{p1.3}
\begin{itemize}
\item[(i)]
\begin{align}\label{1.4}
(A\ltimes B)^\mathrm{T}=B^\mathrm{T}\ltimes A^\mathrm{T}.
\end{align}
\item[(ii)] If $A$ and $B$ are two invertible matrices, then
\begin{align}\label{1.5}
(A\ltimes B)^{-1}=B^{-1}\ltimes A^{-1}.
\end{align}
\end{itemize}
\end{prp}

Hypermatrix \cite{lim13} is a generalized matrix. Roughly speaking, a matrix is a set of data of order $2$, while a hypermatrix is a set of data of order $d>2$. A hypermatrix of order $d$ is closely related to a tensor of covariant order $d$, which is essentially a multilinear (more precisely $d$-th order linear) mapping \cite{boo86}. Meanwhile, it can still be considered as a ``matrix'' in a certain sense, so that some corresponding properties can be studied, such as determinants (now called the hyperdeterminants) \cite{lim13}, eigenvalues and eigenvectors \cite{qi07}, etc.

The theory of compound matrices has found many applications in systems and control theory\cite{bar22,wu22}. The multiplicative compound matrix is defined as follows.

\begin{dfn}\cite{bar22}\label{d1.4} Let $A\in \F^{n\times m}$, $k\leq \min(n,m)$. The $k$-multiplicative compound of $A$, denoted by $A^{(k)}$, is a
$\binom{n}{k}\times \binom{m}{k}$ matrix containing all $k$-minors of $A$ in lexicographical order.
\end{dfn}

\begin{exa}\label{e1.5} Let
$$
A=\begin{bmatrix}
1&2&-1&4\\
-2&0&1&-3\\
3&1&-2&5
\end{bmatrix},
$$
Then
\begin{itemize}
\item[(i)] $A^{(1)}=A.$
\item[(ii)]
$
A^{(2)}=
\begin{bmatrix}
4&-1&5&2&-6&-1\\
-5&1&-7&-3&6&3\\
-2&1&-1&-1&3&-1\\
\end{bmatrix}.
$
\item[(iii)] $A^{(3)}=
\begin{bmatrix}
-1&-3&2&-3\\
\end{bmatrix}.
$
\end{itemize}
\end{exa}

The additive compound matrix is defined as follows.

\begin{dfn} \cite{bar22}\label{d1.6} Let $A\in \F^{n\times n}$, $k\leq n$. The $k$-additive compound of $A$, denoted by $A^{[k]}$, is a
$\binom{n}{k}\times \binom{n}{k}$ matrix defined by:
\begin{align}\label{1.12}
A^{[k]}:=\frac{d}{d\epsilon}\left.\left(I_n+\epsilon A\right)^{(k)}\right|_{\epsilon=0}.
\end{align}
\end{dfn}

\begin{exa}\label{e1.7} Consider
$$
A=\begin{bmatrix}
1&2&-1\\
-2&0&1\\
3&1&-2
\end{bmatrix}
$$
Then
\begin{itemize}
\item[(i)]
$$
\begin{array}{l}
\left(I_3+\epsilon A\right)^{(1)}=I_3+\epsilon A\\
=\begin{bmatrix}
1+\epsilon&2\epsilon&-\epsilon\\
-2\epsilon&1&\epsilon\\
3\epsilon&\epsilon&1-2\epsilon\\
\end{bmatrix}
\end{array}
$$
$$
A^{[1]}=\frac{d}{d\epsilon}\left.(I_3+\epsilon A)\right|_{\epsilon=0}=A.
$$
\item[(ii)]
$$
\begin{array}{l}
\left(I_3+\epsilon A\right)^{(2)}\\
=\begin{bmatrix}
1+\epsilon+4\epsilon^2&\epsilon-\epsilon^2&\epsilon+2\epsilon^2\\
\epsilon-5\epsilon^2&1-\epsilon+\epsilon^2&2\epsilon-3\epsilon^2\\
-3\epsilon-2\epsilon^2&-2\epsilon+\epsilon^2&1-2\epsilon-\epsilon^2\\
\end{bmatrix}
\end{array}
$$
$$
\begin{array}{l}
A^{[2]}=\left.\frac{d}{d\epsilon}\left(I_3+\epsilon A\right)^{(2)}\right|_{\epsilon=0}\\
=\begin{bmatrix}
1&1&1\\
1&-1&2\\
-3&-2&-2\\
\end{bmatrix}
\end{array}
$$
\item[(iii)]
$$
\begin{array}{l}
\left(I_3+\epsilon A\right)^{(3)}\\
=\det\left(I_3+\epsilon A\right)=1-\epsilon+O(\epsilon^2).
\end{array}
$$
$$
A^{[3]}=\frac{d}{d\epsilon}\left.\left(I_3+\epsilon A\right)^{(3)}\right|_{\epsilon=0}=-1.
$$
\end{itemize}
\end{exa}

Some basic properties of compound matrices are listed as follows:

\begin{prp}\cite{bar22}\label{p1.8}
\begin{itemize}
\item[(i)]$A^{(1)}=A.$
\item[(ii)] Assume $A\in \F^{n\times n}$, then $A^{(n)}=\det(A)$.
\item[(iii)] $\left(A^{(k)}\right)^\mathrm{T}=\left(A^\mathrm{T}\right)^{(k)}$.

{$A$ symmetric $\Rightarrow$ $A^{(k)}$ symmetric.}
\end{itemize}
\end{prp}



\begin{thm}\cite{bar22}[Cauchy-Binet Formula]\label{t1.9} 
Let $A\in \F^{n\times m}$,  $B\in \F^{m\times p}$. Fix a positive integer $k\leq \min(n,m,p)$. Then
\begin{align}\label{12.4}
(AB)^{(k)}=A^{(k)}=B^{(k)}.
\end{align}
\end{thm}

\begin{cor}\cite{bar22}\label{c1.10}
\begin{itemize}
\item[(i)]
\begin{align}\label{12.5}
\left(I_n\right)^{(k)}=I_r,\quad r=\binom{n}{k}.
\end{align}
\item[(ii)] If $A\in \F^{n\times n}$ is invertible, then
$A^{(k)}$ is also invertible, and
\begin{align}\label{12.6}
\left(A^{(k)}\right)^{-1}=\left(A^{-1}\right)^{(k)}.
\end{align}
\item[(iii)] $A,B\in \F^{n\times n}$, then
\begin{align}\label{12.601}
\det(AB)=(AB)^{(n)}=A^{(n)}B^{(n)}=\det(A)\det(B).
\end{align}
\item[(iv)] If $A\simeq B$, then $A^{(k)}\simeq B^{(k)}$.
\end{itemize}
\end{cor}

\begin{prp}\cite{bar22}\label{p1.11}
Consider  $A\in \F^{n\times n}$, let $\lambda_i$, $i\in [1,n]$ be the eigenvalues of $A$, and
$v_i$, $i\in [1,n]$ be the corresponding eigenvectors.
Then the eigenvalues of $A^{(k)}$ are
\begin{align}\label{12.7}
\left\{\lambda^{\a}=\prod_{\ell=1}^k\lambda_{i_\ell}\;|\; \a=(i_1,i_2,\cdots,i_{k})\in Q(n,k)\right\}.
\end{align}
Furthermore, if $\a=(i_1,i_2,\cdots,i_k)\in Q(n,k)$ and
$$
W_{\a}:=\left[v_{i_1},v_{i_2},\cdots,v_{i_k}\right]^{(k)}\neq 0,
$$
then $W_{\a}$ is the eigenvector of $A^{(k)}$ corresponding to $\lambda^{\a}$.
\end{prp}

\begin{prp}\cite{bar22}\label{p1.12}
\begin{itemize}
\item[(i)]
\begin{align}\label{13.1}
A^{[k]}=\frac{d}{d\epsilon}\left.\left(\exp(A\epsilon)\right)^{[k]}\right|_{\epsilon=0}.
\end{align}
\item[(ii)]
\begin{align}\label{13.2}
\left(I_n+\epsilon A\right)^{(k)}=I_r+\epsilon A^{[k]} +o(\epsilon).
\end{align}
\item[(iii)]
\begin{align}\label{13.3}
\frac{d}{dt}\left(\exp(At)\right)^{(k)}=A^{[k]} \left(\exp(At)\right)^{(k)}.
\end{align}
\item[(iv)] Let $A,T\in \F_{n\times n}$, with $T$ invertible. Then
\begin{align}\label{13.4}
\left(TAT^{-1}\right)^{[k]}=T^{(k)}A^{[k]} \left(T^{(k)}\right)^{-1}.
\end{align}
\end{itemize}
\end{prp}

\begin{prp}\cite{bar22}\label{p1.13}
Let $A,B\in \C_{n\times n}$. Then
\begin{align}\label{13.5}
\left(A+B\right)^{[k]}=A^{[k]}+B^{[k]}.
\end{align}
\end{prp}

\begin{prp}\cite{bar22}\label{p1.14}
For $A\in \F^{n\times n}$, let $\lambda_i$, $i\in [1,n]$ be the eigenvalues of $A$, and
$v_i$, $i\in [1,n]$ be the corresponding eigenvectors.
Then the eigenvalues of $A^{[k]}$ are
\begin{align}\label{13.6}
\left\{\lambda^{\a}=\dsum_{\ell=1}^k\lambda_{i_\ell}\;|\; \a=(i_1,i_2,\cdots,i_{k})\in Q(n,k)\right\}.
\end{align}
Furthermore, if $\a=(i_1,i_2,\cdots,i_k)\in Q(n,k)$ and
$$
W_{\a}:=\left[v_{i_1},v_{i_2},\cdots,v_{i_k}\right]^{(k)}\neq 0,
$$
then $W_{\a}$ is the eigenvector of $A^{[k]}$ corresponding to $\lambda^{\a}$.
\end{prp}

Finally, we give a list of notations used in this paper.

\begin{enumerate}

\item $\F^n$: $n$ dimensional Euclidean space over $\F$, where $\F$ is a field (Particularly, $\F=\R$ or $\F=\C$).

\item  $\F^{n_1\times n_2\times \cdots \times n_d}$: the $d$-th order hypermatrices of dimensions $n_1,n_2,\cdots,n_d$.

\item $\lcm(n,p)$: the least common multiple of $n$ and $p$.

\item $\d_n^i$: the $i$-th column of the identity matrix $I_n$.

\item $\D_n:=\left\{\d_n^i\vert i=1,\cdots,n\right\}$.

\item $\J_{\ell}:=(\underbrace{1,1,\cdots,1}_{\ell})^\mathrm{T}$.

\item $\cdet$: combinatorial hyperdeterminant.

\item $\ddet$: modified combinatorial hyperdeterminant.

\item $\Det$: slice-based hyperdeterminant.

\item $\ltimes$: semi-tensor product of matrices.

\item $\circledcirc$: semi-tensor product of hypermatrices.

\item $\GL(n^{(d)},\F)$: $n^{(d)}$-general linear group of hypercubics.

\item $\GL(*^{(d)},\F)$: $d$-th order general linear group of hypercubics.
\end{enumerate}

The purpose of this paper is twofold: 
\begin{itemize}
\item[(i)] To generalize the STP of matrices to the STP of hypermatrices, that is, to define a product of two hypermatrices of arbitrary orders and arbitrary dimensions, and to show that some of the above properties can be extended to the STP of hypermatrices; (2) The determinant of matrices is also extended to several types of hyperdeterminants of hypermatrices.
The monoid (semigroup with identity) and group structures of matrices are extended to the monoid and group structures of hypermatrices. Finally, the general linear group of hypermatrices is also introduced as a Lie group.
\item[(ii)] Two types of compound hypermatrices are proposed: multiplicative and additive. Using the semi-tensor product of hypermatrices, it is shown that most of the properties of compound matrices can be extended to compound hypermatrices.
\end{itemize}


\section{Matrix Expression of Hypermatrices}

\begin{dfn}\label{d2.1}
\begin{itemize}
\item[(i)] A set of order $d$ data
\begin{align}\nonumber
A:=\{a_{i_1,i_2,\cdots,i_d}\;|\; i_s\in[1,n_s],\; s\in[1,d]\}\\\label{2.1}
\in \F^{n_1\times \cdots \times n_d}
\end{align}
is called an $d$-th order hypermatrix ($d$-hypermatrix for short) of dimensions $n_1\times n_2\times \cdots \times n_d$. The set of  $d$-hypermatrix of dimension $n_1\times n_2\times \cdots \times n_d$ is denoted by  $\F^{n_1\times n_2\times\cdots\times n_d}$, where $a_{i_1,i_2,\cdots,i_d}\in \F$ and $\F$ can be $\R$ or $\C$ (or any other field).
\item[(ii)] $A\in \F^{\overbrace{n\times n\times\cdots\times n}^d}$ is called a $d$-hypercubic.
\item[(iii)] $A\in \F^{n_1\times n_2\times\cdots\times n_d}$ with $n_1=n_d$ is called a $d$-hypersquare.
\end{itemize}
\end{dfn}

The elements of $A$ can be arranged into a matrix by a particular partition of indices, called the matrix expression of $A$.

\begin{dfn}\label{d2.2} Let
\begin{align}\label{2.2}
A=[a_{i_1,i_2,\cdots,i_d}]\in \F^{n_1\times n_2\times\cdots\times n_d},
\end{align}
$\a=\{\a_1,\a_2,\cdots,\a_s\}\subset \{1,2,\cdots,d\}$ and  $\b=\{\b_1,\b_2,\cdots,\b_t\}\subset \{1,2,\cdots,d\}$ and
\begin{align}\label{2.3}
\a \bigcup \b =\{1,2,\cdots,d\}
\end{align}
be a partition, where $s+t=d$. Then
\begin{align}\label{2.4}
M^{\a}_A\in \F^{n_{\a}\times n_{\b}}
\end{align}
is a matrix of dimension
$$
n_{\a}:=\prod_{i=1}^sn_{\a_i};\quad n_{\b}:=\prod_{j=1}^tn_{\b_j},
$$
where its rows are arranged by the index arrangement
$$\mathrm{Id}(\a_1,\a_2,\cdots,\a_s;n_{\a_1},n_{\a_2},\cdots,n_{\a_s})\in Q(s,d),
$$ and
its columns are arranged by the index
$$\mathrm{Id}(\b_1,\b_2,\cdots,\b_t;n_{\b_1},n_{\b_2},\cdots,n_{\b_t})\in Q(t,d),
$$
where $Q(m,n)$ is the set of $m$ sub-indices of the set of $n$ indices.
$M^{\a}_A$ is called the matrix expression of $A$ with respect to the partition $\{\a,\b=\a^c\}$.
\end{dfn}

\begin{rem}\label{r2.3}
\begin{itemize}
\item[(i)] For simplicity, we always assume $\a_1<\a_2<\cdots<\a_s$,  $\b_1<\b_2<\cdots<\b_s$.
\item[(ii)] Index arrangement $\mathrm{Id}$ means the indexed data are arranged in alphabetic order \cite{che12}.
\end{itemize}
\end{rem}

\begin{exa}\label{e2.4}
Given $A=[a_{i_1,i_2,i_3}]\in \F^{2\times 3\times 2}$. Then
\begin{itemize}
\item[(i)]
$$
\begin{array}{ccl}
M_A^{\emptyset}&=&[a_{111},a_{112},a_{121},a_{122},a_{131},a_{132},\\
~&~&a_{211},a_{212},a_{221},a_{222},a_{231},a_{232}].
\end{array}
$$
\item[(ii)]
$$
M_A^{\{1\}}=
\begin{bmatrix}
a_{111}&a_{112}&a_{121}&a_{122}&a_{131}&a_{132}\\
a_{211}&a_{212}&a_{221}&a_{222}&a_{231}&a_{232}
\end{bmatrix};
$$
$$
M_A^{\{2\}}=
\begin{bmatrix}
a_{111}&a_{112}&a_{211}&a_{212}\\
a_{121}&a_{122}&a_{221}&a_{222}\\
a_{131}&a_{132}&a_{231}&a_{232}
\end{bmatrix}; \quad \mbox{etc.}
$$
\item[(iii)]
$$
M_A^{\{1,2\}}=
\begin{bmatrix}
a_{111}&a_{112}\\
a_{121}&a_{122}\\
a_{131}&a_{132}\\
a_{211}&a_{212}\\
a_{221}&a_{222}\\
a_{231}&a_{232}
\end{bmatrix};
$$
$$
M_A^{\{1,3\}}=
\begin{bmatrix}
a_{111}&a_{121}&a_{131}\\
a_{112}&a_{122}&a_{132}\\
a_{211}&a_{221}&a_{231}\\
a_{212}&a_{222}&a_{232}
\end{bmatrix}; \quad \mbox{etc.}
$$
\item[(iv)]
$$
M_A^{\{1,2,3\}}=\left(M_A^{\emptyset}\right)^\mathrm{T}.
$$
\end{itemize}
\end{exa}

Denote
$$
V_A:=M_A^{\emptyset};\quad M_A:=M_A^{\{1\}}.
$$

In the sequel, $M_A$ plays a particularly important role. Let $A\in \F^{n_1\times n_2\times \cdots\times n_d}$. Then
\begin{align}\label{2.401}
M_A=[A_1,A_2,\cdots,A_q],
\end{align}
where $A_i\in \F^{n_1\times n_d}$, $i\in [1,q]$, ($q=\prod_{i=2}^{d-1}n_i$) are called slices of $A$.

\begin{prp}\label{p2.5} Given $A\in \F^{n_1\times n_2\times \cdots \times n_d}$,  $\a\in Q(s,d)$ and $\b\in Q(t,d)$ ($s+t=d$) be a partition of $[1,d]$. Then $A$ can be considered as a multi-linear mapping $\pi_A^{(\a)}:\F^{n}\ra \F^{m}$, where $m=\prod_{i\in \a}n_i$ and $n=\prod_{i\in \b}n_i$, and
\begin{align}\label{2.5}
\pi_A^{\a}(x):=M^{\a}_Ax\in \F^m,\quad x\in \F^n.
\end{align}
\end{prp}

\begin{dfn}\label{d2.6} \cite{lim13}
\begin{itemize}
\item[(i)] Given a $d$-hypermatrix $A=\left[ a_{j_1,j_2,\cdots,j_d} \right]\in \F^{n_1\times n_2\times \cdots\times n_d}$, and assume $\sigma\in {\bf S}_d$.
The $\sigma$-transpose of $A$ is
\begin{align}\label{2.6}
A^{\sigma}:=\left[ a_{j_{\sigma(1)}\cdots j_{\sigma(d)}}\right]\in \F^{n_{\sigma(1)}\times \cdots\times n_{\sigma(d)}}.
\end{align}
\item[(ii)] A $d$-hypercubic $A\in \overbrace{\F^n\times \cdots\times \F^n}^d$ is called symmetric if $\forall \sigma\in {\bf S}_d$, $A^{\sigma}=A$; $A$ is called skew-symmetric if $\forall \sigma\in {\bf S}_d$, $A^{\sigma}=\sgn(\sigma)A$.
\end{itemize}
\end{dfn}

\begin{prp}\label{p2.7} A
 $2$-hypercubic $A\in \F^{n\times n}$ is (skew-)symmetric, if and only if,
 $M_A$ is (skew-)symmetric.
\end{prp}

\begin{rem}\label{r2.8} The above arguments stand true even when $\F$ is the set of perfect hypercomplex numbers (PHNs) \cite{che21b}. In fact, most of arguments throughout this paper also hold for PHNs.
\end{rem}

\section{Semi-tensor Product of Hypermatrices}

\begin{dfn}\label{d3.1} Let $A\in \F^{m\times s\times n}$ and $B\in \F^{p\times s\times q}$ be two $3$-hypermatrices, $t=\lcm(n,p)$.
Then
$$
\begin{array}{l}
M_A=\left[A_1,A_2,\cdots,A_{s}\right],\\
M_B=\left[B_1,B_2,\cdots,B_{s}\right],\\
\end{array}
$$
where
$$
A_i\in \F^{m\times n},\quad B_i\in \F^{p\times q},\quad i\in[1,s].
$$
Then
the SPTH of $A$ and $B$ is defined by
\begin{align}\label{3.1}
A\circledcirc B:=C,
\end{align}
where
$$
M_C=\left[C_1,C_2,\cdots,C_{s}\right]\in \F^{(mt/n)\times s \times (qt/p)},
$$
and
$$
C_i=A_i\ltimes B_i,\quad i\in[1,s].
$$
\end{dfn}

\begin{exa}\label{e3.2}
Given $A\in \F^{2\times 2\times 3}$ and $B\in \F^{2\times 2\times 2}$ with
$$
M_A=\begin{bmatrix}
a_{111}&a_{112}&a_{113}&a_{121}&a_{122}&a_{123}\\
a_{211}&a_{212}&a_{213}&a_{221}&a_{222}&a_{223}\\
\end{bmatrix}
:=[A_1,A_2],
$$
$$
M_B=\begin{bmatrix}
b_{111}&b_{112}&b_{121}&b_{122}\\
b_{211}&b_{212}&b_{221}&a_{222}\\
\end{bmatrix}:=[B_1,B_2].
$$
Let
$$
C=A\circledcirc B\in \F^{4\times 2\times 6}.
$$
Then
$$
M_C=[A_1\ltimes B_1, A_2\ltimes B_2].
$$
\end{exa}

Next, we extend the STPH to general cases:

\begin{itemize}
\item Case 1, $d>3$:
\end{itemize}

Assume $A\in \F^{n_1\times n_2\times \cdots \times n_d}$. Denote $s=\prod_{i=2}^{d-1}n_i$, and define
$\pi: \F^{n_1\times n_2\times \cdots \times n_d}\ra \F^{n_1\times s \times n_d}$ by
\begin{align}\label{3.2}
\pi:~a_{i_1,i_2,\cdots,i_d}\mapsto b_{i_1,t,i_d},
\end{align}
where
\begin{align}\label{3.3}
\begin{array}{ccl}
t&=&(i_2-1)n_3n_4\cdots n_{d-1}+ (i_3-1)n_4n_5\cdots n_{d-1}+\\
~&~&\cdots+(i_{d-2}-1)n_{d-1}+i_{d-1}.\\
\end{array}
\end{align}
Then it is easy to see that $\pi$ is a bijective mapping. Hence we can extend the STPH defined in Definition \ref{d3.1} to more general cases.

\begin{dfn}\label{d3.3}
Assume $A\in \F^{m\times s_2\times \cdots \times s_{d-1}\times n}$ and $B\in \F^{p\times s_2\times \cdots \times s_{d-1}\times q}$, then
\begin{align}\label{7.3}
A\circledcirc B:=\pi^{-1}\left(\pi(A) \circledcirc \pi(B)\right)\in \F^{(mt/n)\times s_2\times \cdots \times s_{d-1}\times (qt/p)}.
\end{align}
\end{dfn}

\begin{itemize}
\item Case 2, $d=3$ and $s_1\neq s_2$:
\end{itemize}

\begin{dfn}\label{d3.4}
Assume $A\in \F^{m\times s_1\times n}$ and $B\in \F^{p\times s_2\times q}$ and $\lcm(s_1,s_2)=s$,
construct
$$
{\begin{array}{l}
M_{\tilde{A}}:=\J^\mathrm{T}_{s/s_1}\otimes M_A;\\
M_{\tilde{B}}:=\J^\mathrm{T}_{s/s_2}\otimes M_B.\\
\end{array}}
$$
Then $\tilde{A}\in \F^{m\times s\times n}$ and $\tilde{B}\in \F^{p\times s\times q}$.
Define
\begin{align}\label{7.4}
A\circledcirc B:=\tilde{A}\circledcirc \tilde{B}\in \F^{(mt/n)\times s\times (qt/p)}.
\end{align}
\end{dfn}

Combining Definitions \ref{d3.1}, \ref{d3.3}, and \ref{d3.4}, one sees easily that the STPH of two arbitrary hypermatrices is properly defined.
Hence it is enough to consider the case of Definition  \ref{d3.1}.

\begin{exa}\label{e3.5}
Given $A\in \F^{m\times 3\times n}$ and $B\in \F^{p\times 2\times 2\times q}$. Express
$$
\begin{array}{l}
M_A=\left[A_1,A_2,A_3\right],\\
M_B=\left[B_{11},B_{12},B_{21},B_{22}\right].\\
\end{array}
$$
Then $C=A\circledcirc B$ can be calculated by
$$
{\begin{array}{ccl}
M_C&=&[\J_4^\mathrm{T} \otimes M_A ] \circledcirc [\J_3^\mathrm{T} \otimes M_B]\\
~&=&[A_1\ltimes B_{11}, A_2\ltimes B_{12}, A_3\ltimes B_{21}, A_1\ltimes B_{22}\\
~&~&~A_2\ltimes B_{11}, A_3\ltimes B_{12},A_1\ltimes B_{21},  A_2\ltimes B_{22}\\
~&~&~A_3\ltimes B_{11}, A_1\ltimes B_{12},A_2\ltimes B_{21},  A_3\ltimes B_{22}].
\end{array}}
$$
\end{exa}

Next, we show some basic properties of STPH.

Denote by
$$
\F^{\infty^{\infty}}=\dsum_{d=1}^{\infty}\dsum_{n_1=1}^{\infty}\cdots\dsum_{n_d=1}^{\infty}\F^{n_1\times n_2\times \cdots \times n_d}.
$$

To include $\F$, $\F^{n_1\times n_2}$, we consider
$$
a=a (1\times 1\times 1);
$$
and
$A\in \F^{n_1\times n_2}$
as
$$
A\in \F^{n_1\times 1\times n_2}.
$$
Then the STP of hypermatrices is also applicable to $\F$, $\F^{n_1\times n_2}$.

\begin{dfn}\label{d5.1}
Let $a\in \F$ and $B\in \F^{p\times s\times q}$. Then
\begin{align}\label{5.1}
a\circledcirc B:=aB.
\end{align}
Let $A\in \F^{m\times n}$ and $B\in \F^{p\times s\times q}$, and $t=lcm(n,p)$. Then
\begin{align}\label{5.2}
A\circledcirc B:=C\in \F^{mt/n\times s\times qt/p},
\end{align}
where
$$
M_C=\left[A\ltimes B_1,A\ltimes B_2,\cdots,A\ltimes B_s\right].
$$
\end{dfn}

\begin{prp}\label{p5.2} Assume $A,B,C\in  \F^{\infty^{\infty}}$, then
\begin{align}\label{5.3}
A\circledcirc (B\circledcirc  C)=(A\circledcirc B)\circledcirc  C.
\end{align}
\end{prp}

\begin{prp}\label{p5.3} Assume $A,B\in \F^{n_1\times n_2\cdots\times n_d}$ and $C\in  \F^{\infty^{\infty}}$, then
\begin{align}\label{5.4}
\begin{array}{l}
(A+B)\circledcirc  C=A\circledcirc C+B\circledcirc C,\\
C \circledcirc  (A+B)=C\circledcirc A+C\circledcirc B.
\end{array}
\end{align}
\end{prp}

\begin{prp}\label{p5.4} Assume $A$ and $B$ are two invertible hypersquares, then
\begin{align}\label{5.5}
(A\circledcirc B)^{-1}=B^{-1}\circledcirc A^{-1}.
\end{align}
\end{prp}

\section{Hyperdeterminants}

\begin{dfn}\label{d4.1} \cite{lim13} Let $A\in \F^{\overbrace{n\times \cdots\times n}^d}$ be a $d$-hypercubic. The combinatorial hyperdeterminant (CH-determinant) is defined by
\begin{align}\label{4.1}
\begin{array}{l}
\cdet(A)=
\frac{1}{n!}\dsum_{\sigma_1,\cdots, \sigma_{d}\in {\bf S}_n}\prod\limits_{j=1}^{d}
\sgn(\sigma_j)\prod\limits_{i=1}^n a_{\sigma_1(i),\cdots,\sigma_{d}(i)}.
\end{array}
\end{align}
\end{dfn}

\begin{rem}  Combinatorial hyperdeterminant has some nice properties. Unfortunately, for an odd order $d$, the combinatorial hyperdeterminent of a $d$-hypercubic is identically zero \cite{lim13}. So it is not suitable for our approach where, mostly, $d=3$. So we provide a modification as follows.
\end{rem}

\begin{dfn}\label{d4.2}
Let $A$ be a $d$-hypercubic with dimension $n_i=n$, $i\in[1,d]$. The modified combinatorial hyperdeterminant (MCH-determinant) of $A$ is defined by
\begin{align}\label{4.2}
\begin{array}{l}
\ddet(A)=\\
\dsum_{\sigma_1,\cdots,\sigma_{d-1}\in {\bf S}_n}\prod\limits_{j=1}^{d-1}
\sgn(\sigma_j)\prod\limits_{i=1}^n a_{i,\sigma_1(i),\cdots,\sigma_{d-1}(i)}.
\end{array}
\end{align}
\end{dfn}

\begin{prp}\label{p4.201}
Assume $d=2$, then
\begin{align}\label{4.3}
\ddet(A)=\det(M_A).
\end{align}
\end{prp}

\begin{prp}\label{p4.202}
When $d$ is even
\begin{align}\label{4.301}
\ddet(A)=\cdet(A).
\end{align}
\end{prp}

The following definition is more suitable for our purpose:

\begin{dfn}\label{d4.3}
\begin{itemize}
\item[(i)] Let $A$ be a $d$-hypersquare with dimension $n_1=n_d=n$, $d\geq 3$. Denote $s=\prod\limits_{i=2}^{d-1}n_i$, and
$$
M_A:=[A_1,A_2,\cdots,A_s].
$$
Then the slice-based hyperdeterminant (SH-determinant) of $A$ is defined by
\begin{align}\label{4.4}
\Det(A):=\prod\limits_{i=1}^{s}\det(A_i).
\end{align}
\item[(ii)]  $d$-hypersquare $A$ is called non-singular (invertible), if $$\Det(A)\neq 0.$$
\item[(iii)]
The  $d$-hypersquare $B$ is called the inverse of $A$, if
$$
M_B:=[A^{-1}_1,A^{-1}_2,\cdots,A^{-1}_s].
$$
\end{itemize}
\end{dfn}

\begin{exa}\label{e4.4} Assume $A\in \F^{2\times 2\times 2}$ with
\begin{align}\label{4.5}
M_A=\begin{bmatrix}
a_{111}&a_{112}&a_{121}&a_{122}\\
a_{211}&a_{212}&a_{221}&a_{222}\\
\end{bmatrix}
\end{align}

\begin{itemize}
\item[(i)]

$$
\begin{array}{rl}
~&\cdet(A)\\
=&\frac{1}{2}\dsum_{\sigma_1,\sigma_2,\sigma_3=1}^2a_{\sigma_1(1)\sigma_2(1)\sigma_3(1)}\times a_{\sigma_1(2)\sigma_2(2)\sigma_3(2)}\\
=&\frac{1}{2}(a_{111}a_{222}-a_{112}a_{221}-a_{121}a_{212}+a_{122}a_{211}\\
~&-a_{211}a_{122}+a_{212}a_{121}+a_{221}a_{112}-a_{222}a_{111})\\
=&0.
\end{array}
$$
\item[(ii)]
$$
\begin{array}{ccc}
\ddet(A)=\dsum_{\sigma_1=1}^2\dsum_{\sigma_2=1}^2a_{1\sigma_1(1)\sigma_2(1)}a_{2\sigma_1(2)\sigma_2(2)}\\
=a_{111}a_{222}-a_{112}a_{221}-a_{121}a_{212}+a_{122}a_{211}.
\end{array}
$$
\item[(iii)]
$$
\begin{array}{ccc}
\Det(A)=\det(A_1)\det(A_2)\\
=\left(a_{111}a_{212}-a_{112}a_{211}\right)\left(a_{121}a_{222}-a_{122}a_{211}\right).
\end{array}
$$
\end{itemize}
\end{exa}

\begin{dfn}\label{d4.5}
\begin{itemize}
\item[(i)] Let $A\in \F^{m\times n}$ and $m\leq n$. Denote by
$A^{(m)}=(a_1,a_2,\cdots,a_r)$, where $r=\binom{n}{m}$. Then
\begin{align}\label{4.6}
\Det(A):=\prod_{i=1}^ra_i.
\end{align}
\item[(ii)] Let $A\in \F^{n_1\times n_2\times \cdots \times n_d}$. Denote by
$M_A=[A_1,A_2,\cdots,A_s]$, where $s=\prod_{i=2}^{d-1}n_i$. 
Then
\begin{align}\label{4.7}
\Det(A):=\prod_{i=1}^s\Det(A_i).
\end{align}
\end{itemize}
\end{dfn}

Next, with the newly introduced notions of STPH and hyperdeterminant, we reveal some basic algebraic structure of hypermatrices.

\begin{prp}\label{p6.1} $\left(\F^{\infty^{\infty}},\circledcirc \right)$ is a monoid (semi-group with identity), with identity $1\in \F$.
\end{prp}

Denote $J_n^s\in \F^{n\times s\times n}$ as
$$
M_{J_n^s}=\underbrace{[I_n,I_n,\cdots,I_n]}_s.
$$

\begin{dfn}\label{d6.2}  Let $\F^{n\times s\times n}$ be the set of hypersquare, where $s=(s_2,s_3,\cdots,s_{d-1})$, $d\geq 2$. Consider
$$
G=\left\{A\in \F^{n\times s\times n}\;|\; \Det(A)\neq 0\right\}.
$$
Denote by
\begin{align}\label{6.1}
\GL(n^s,\F):=\left(G,\circledcirc\right).
\end{align}

Then $\GL(n^s,\F)$ is a group  with identity
$J^s_n$, called the general linear group of hypermatrices.
\end{dfn}

\begin{prp}\label{p6.2}
$\GL(n^s,\F)$ is a Lie-group with natural manifold structure.
\end{prp}

\begin{dfn}\label{d6.3}
Define
\begin{align}\label{6.2}
\GL(\infty^{\infty},\F):=\bigcup_{n=1}^{\infty}\bigcup_{d=1}^{\infty}\GL(n^s,\F).
\end{align}
Then $\GL(\infty^{\infty},\F)$ is called the dimension-free general linear group of hypermatrices.
\end{dfn}

\begin{prp}\label{p6.4}
$\GL(\infty^{\infty},\F)$ is a dimension-free Lie-group with dimension-free manifold structure \cite{chepr}.
\end{prp}

\section{Compound Hypermatrices}

The extension of compound matrices to compound hypermatrices is straightforward:

\begin{dfn}\label{d7.1} Let $A\in \F^{n_1,n_2,\cdots,n_d}$, $k\leq \min(n_1,n_d)$. Denote
$$
M_A:=[A_1,A_2,\cdots,A_s],
$$
where $s=\prod\limits_{i=2}^{d-1}n_i$.
\begin{itemize}
\item[(i)]  The $k$-multiplicative  compound hypermatrix of $A$, denoted by $A^{(k)}$, is defined by
\begin{align}\label{7.1}
M_{A^{(k)}}:=[A_1^{(k)},A_2^{(k)},\cdots,A_s^{(k)}].
\end{align}
\item[(ii)]  The $k$-additive  compound hypermatrix of $A$, denoted by $A^{[k]}$, is defined by
\begin{align}\label{7.2}
M_{A^{[k]}}:=[A_1^{[k]},A_2^{[k]},\cdots,A_s^{[k]}].
\end{align}
\end{itemize}
\end{dfn}

The following properties are a direct consequence of Definition \ref{d7.1} and the corresponding properties of compound matrices.

\begin{prp}\label{p7.2}
\begin{itemize}
\item[(i)]
\begin{align}\label{7.3}
A^{(1)}=A.
\end{align}
\item[(ii)] Assume $A$ is a hypersquare, that is, $A\in \F^{n_1\times n_2\times \cdots \times n_{d}}$ and $n_1=n_d$, then
\begin{align}\label{7.4}
A^{(n)}=[\det(A_1),\det(A_2),\cdots,\det(A_s)].
\end{align}
\end{itemize}
\end{prp}

\begin{thm}\label{t7.4}(Cauchy-Binet Formula)
Let $A\in \F^{n_1\times n_2\times \cdots\times n_{d}}$, $B\in \F^{m_1\times m_2\times \cdots\times m_{d}}$, where $n_d=m_1$ and $n_i=m_i$, $i\in [2,d-1]$.  Fix a positive integer $k\leq \min(n_1,n_d,m_1)$, then
\begin{align}\label{7.5}
(A\circledcirc B)^{(k)}=A^{(k)}\circledcirc B^{(k)}.
\end{align}
\end{thm}

\begin{rem}\label{r7.5} When $A\in \F^{m\times n}$, $B\in \F^{n\times r}$.  Fix a positive integer $k\leq \min(m,n,r)$, then
\begin{align}\label{7.6}
(AB)^{(k)}=A^{(k)}B^{(k)}.
\end{align}
This is the classical Cauchy-Binet Formula.
\end{rem}

\begin{cor}\label{c7.6}
\begin{itemize}
\item[(i)]
\begin{align}\label{7.7}
\left(J_n^d\right)^{(k)}=\left(J_r^d\right)^{(k)}, \quad r=\binom{n}{k}.
\end{align}
Particularly,
\begin{align}\label{7.8}
\left(I_n\right)^{(k)}=I_r, \quad r=\binom{n}{k}.
\end{align}
\item[(ii)] If $A\in \F_{n_1\times n_2\times \cdots \times n_d}$, $n_1=n_d=n$, is invertible, then
$A^{(k)}$ is also invertible, and
\begin{align}\label{7.9}
\left(A^{(k)}\right)^{-1}=\left(A^{-1}\right)^{(k)}.
\end{align}
\end{itemize}
\end{cor}

\begin{dfn}\label{d7.7}
Let $A,B\in \F_{n_1\times n_2\times \cdots \times n_d}$, $n_1=n_d=n$. $A$ and $B$ are said to be similar, denoted by $A\simeq B$, if
there exists a nonsingular $T\in \F_{n_1\times n_2\times \cdots \times n_d}$
 such that
\begin{align}\label{7.10}
T^{-1}\circledcirc A\circledcirc  T=B.
\end{align}
\end{dfn}

\begin{prp}\label{p7.8} Let $A,B\in \F^{n_1\times n_2\times \cdots \times n_d}$, $n_1=n_d=n$. If $A \simeq B$, then
\begin{itemize}
\item[(i)] \begin{align}\label{7.11}
A^{(k)}\sim B^{(k)},\quad k\leq n;
\end{align}
\item[(ii)] \begin{align}\label{7.1101}
A^{[k]}\sim B^{[k]},\quad k\leq n;
\end{align}
\end{itemize}
\end{prp}

\begin{dfn}\label{d7.9} Assume $A\in \F^{n_1\times n_2\times \cdots \times n_d}$, $n_1=n_d=n$, and $X\in \F^{n_1\times n_2\times \cdots \times n_{d-1}\times 1}$. Moreover,
$$
M_X=[X_1,X_2,\cdots,X_s], \quad s=\prod_{i=2}^{d-1}n_i,
$$
and $X_i\neq 0$, $i\in [1,s]$. If there exists $\lambda=(\lambda_1,\cdots,\lambda_s)\in \F^{s}$ such that
\begin{align}\label{7.12}
A\circledcirc X=\lambda\circledcirc X,
\end{align}
then $\lambda$ is called the eigenvalue of $A$ with eigenvector $X$.
\end{dfn}

Using Definition \ref{d7.9}, Propositions \ref{p1.11} and \ref{p1.14} can be extended to compound hypermatrices easily.

\section{Conclusion}

In this paper the STP of matrices was extended to the STP of two arbitrary hypermatrices. We showed that almost fundamental properties of the STP of matrices can be extended to that of hypermatrices. Three determinants of hypermatrices, namely the CH-determinant, the MCH-determinant and the SH-determinant, are introduced and studied. The monoid and the group of hypermatrices are also introduced. Then the general linear group of hypermatrices, as a Lie group, is introduced and studied. Finally, the compound hypermatrix is presented and some interesting properties are revealed.


\begin{thebibliography}{00}
\bibitem{bar22}  E. Bar-Shalom, O. Dalin, M. Margaliot, Compound matrices in systems and control theory: a tutorial, {\it arXiv:2204.00676v1},  2022.
%
\bibitem{boo86} W.M. Boothby, {\it Introduction to Differentiable Manifolds and Riemannian Geometry}, 2nd Ed., Elsevier, London, 1986.
%
\bibitem{che11} D. Cheng, H. Qi, Z. Li, {\it Analysis and Control of Boolean Networks - A Semi-tensor Product
Approach}, Springer, London, 2011.
\bibitem{che12}  D. Cheng, H. Qi, Y. Zhao, {\it An Introduction to Semi-tensor Product of Matrices and Its Applications}, World Scientific, Singapo, 2012.
%
\bibitem{che19} D. Cheng, On equivalence of matrices, {\it Asian Journal of Math.}, Vol. 23, No. 2, 257-348, 2019.
%
\bibitem{che19b} D. Cheng, {\it From Dimension-Free Matrix Theory to Cross-Dimensional Dynamic Systems}, Elsevier, London, 2019.
%
\bibitem{che21} D. Cheng, Y. Wu, G. Zhao, S. Fu, A comprehensive survey on STP approach to finite games, {\it J. Sys. Sci. Compl.}, Vol. 34, No. 5, 1666-1680, 2021.

\bibitem{che21b}  D. Cheng, Z. Ji, J. Feng, S. Fu, J. Zhao, Perfect hypercomplex algebras: Semi-tensor product approach, {\it Math. Model. Contr.}, Vol. 1, No. 4, 177-187, 2021.
%
\bibitem{chepr} D. Cheng, From dimension-free manifold to dimension-varying control system, http:arxiv.org/abs/2206.04461, 2023.
%
\bibitem{for16} E. Fornasini, M.E. Valcher, Recent developments in Boolean networks control, {\it J. Contr. Dec.}, Vol. 3, No. 1, 1-18, 2016.
%
\bibitem{li18} H. Li, G. Zhao, M. Meng, J. Feng, A survey on applications of semi-tensor product method in engineering, {\it Science China}, Vol. 61, 010202:1-010202:17, 2018.
%
\bibitem{lim13}
L. Lim, Tensors and Hypermatrices, in L. Hogben (Ed.) {\it Handbook of Linear Algebra} (2nd ed.), Chapter 15,
 Chapman and Hall/CRC.https://doi.org/10.1201/b16113, 2013.
%
\bibitem{lu17}  J. Lu, H. Li, Y. Liu, F. Li, Survey on semi-tensor product method with its applications in logical networks and other finite-valued systems, {\it IET Contr. Theory Appl.}, Vol. 11, No. 13, 2040-2047, 2017.
%
\bibitem{muh16} A. Muhammad, A. Rushdi, F.A. M. Ghaleb, A tutorial exposition of semi-tensor products of matrices with a stress on their representation of Boolean function, {\it JKAU Comp. Sci.}, Vol. 5, 3-30, 2016.
%
\bibitem{qi07} L. Qi, Eigenvalues and invariants of tensors, {\it J. Math. Anal. Appl.}, Vol. 325, 1363-1377, 2007.
%
\bibitem{wu22} C. Wu, R. Pines, M. Margaliot, J.J. Slotine, Generalization of the multiplicative and additive compounds of square matrices and contraction theory in the hausdorff dimension, {\it IEEE Trans. Aut. Contr.}, Vol. 67, No. 9, 4629-4644, 2022.

\bibitem{yan22} Y. Yan, D. Cheng, J. Feng, H. Li, J. Yue, Survey on applications of algebraic state space theory of logical systems to finite state machines, {\it Sci. China Inf. Sci.}, https://doi.org/10.1007/s11432-22-3538-4, 2022.
\end{thebibliography}
\end{document}